\documentclass[aps,prl,reprint,superscriptaddress,amsmath,amssymb,amsfonts]{revtex4-2}

\usepackage{graphicx}
\usepackage{dcolumn}
\usepackage{bm}
\usepackage{hyperref} 
\usepackage[mathlines]{lineno}
\usepackage{xcolor}

\begin{document}


\title{Attosecond control of solid-state high harmonic generation using $\omega$-$3\omega$ fields}


\author{Adam Gindl}
    \affiliation{Faculty of Mathematics and Physics, Charles University, Ke Karlovu 3, 12116 Prague 2, Czech Republic}
\author{Pawan Suthar}
    \affiliation{Faculty of Mathematics and Physics, Charles University, Ke Karlovu 3, 12116 Prague 2, Czech Republic}
\author{František Trojánek}
    \affiliation{Faculty of Mathematics and Physics, Charles University, Ke Karlovu 3, 12116 Prague 2, Czech Republic}
\author{Petr Malý}
    \affiliation{Faculty of Mathematics and Physics, Charles University, Ke Karlovu 3, 12116 Prague 2, Czech Republic}
\author{Thibault J.-Y. Derrien}
    \affiliation{HiLASE Centre, Institute of Physics, Academy of Sciences of the Czech Republic, 25241 Dolní Břežany, Czech Republic}
    \affiliation{Institute of Physics of the Czech Academy of Sciences, Na Slovance 1999/2, 18200 Prague, Czech Republic}
   \affiliation{IT4Innovations, Technical University of Ostrava, 17. listopadu 2172/15, 70800 Ostrava-Poruba, Czech Republic} 
\author{Martin Kozák}
    \email{m.kozak@matfyz.cuni.cz}
\affiliation{Faculty of Mathematics and Physics, Charles University, Ke Karlovu 3, 12116 Prague 2, Czech Republic}
\thanks{Corresponding author: m.kozak@matfyz.cuni.cz}

\begin{abstract}
High harmonic spectra generated in condensed matter carry the fingerprints of sub-cycle electronic motion and the energy structure of the studied system. Here we show that tailoring the waveform of mid-infrared driving light by using a coherent combination with its third harmonic frequency allows to control the time of electron tunneling to the conduction band within each half-cycle of the fundamental wave with attosecond precision. We introduce an experimental scheme in which we simultaneously monitor the modulation of amplitude and emission delays of high harmonic radiation and the excited electron population generated in crystalline silicon as a function of the relative phase between the $\omega$-3$\omega$ fields. We observe that the mutual $\omega$-3$\omega$ phase required for the maximum yield of high harmonic generation is shifted by approximately $\pi/2$ with respect to the phase leading to maximal generated carrier population. The observed emission delays of high harmonic photons of up to few hundred attoseconds scale with the time delay and with the ratio between the electric field amplitudes of the two-color fields. These results reveal the connection between electron tunneling and high harmonic emission processes in solids.
\end{abstract}

\maketitle

Coherent nonlinear electron dynamics driven by strong oscillating fields of nonresonant electromagnetic waves in atomic or solid-state systems leads to generation of high energy photons \cite{McPherson1987,Ferray1988}. High harmonic generation (HHG) has opened the door to attosecond physics and spectroscopic techniques based on HHG have recently become indispensable tools for investigating the ultrafast electron dynamics in various physical systems \cite{Corkum2007,Krausz2009,Levesque2007,Ghimire2011,Luu2015}. HHG in condensed matter is a consequence of coherent nonlinear interband polarization and intraband electron currents \cite{Ghimire2019}. The interband source of high energy photons can be understood in the framework of a three-step semi-classical model adopted from atomic physics \cite{Corkum1993}. The electron first tunnels to the conduction band in a narrow time window close to each maximum and minimum of the oscillating electric field of the driving wave. Subsequently, the coherent electron-hole wavepacket is accelerated by the laser field in the crystal and may eventually recombine while emitting high energy photon. On the other hand, the intraband contribution to HHG is a consequence of anharmonic motion of electrons in nonparabolic bands. As a result, the amplitude, phase and polarization of the emitted high harmonic radiation are tightly linked to the band structure of the material and to the polarization state and time profile of the electric field waveform of the driving pulse \cite{You2017,Yoshikawa2017,Klemke2019,You2017CEP,Schubert2014,Garg2016,Watanabe1994,Dudovich2006,Vampa2015,VampaPRL2015,Orenstein2019,Mitra2020,Uzan2020,Severt2021,Langer2016}. In extreme cases, the nonperturbative sub-cycle electron dynamics in strong low-frequency fields may also lead to dynamical Bloch oscillations \cite{Schubert2014}  and Wannier-Stark localization \cite{Schiffrin2013,Schmidt2018}.

Although high harmonic spectroscopy in condensed matter has become a widely used technique, there are only few experiments separating the individual processes contributing to the emission of the coherent high energy photons \cite{Langer2016,Garg2016}. These include coherent two-color optical control, which has initially been developed to steer electron currents in semiconductors via quantum path interference by the $\omega$-2$\omega$ field superposition \cite{Dupont1995,Hache1997}. Since then, the two-color coherent control has been demonstrated in many physical systems, including the control of quantum wave function of electrons in atoms \cite{Weinacht1999}, electron photoemission from metals \cite{Forster2016,Li2021}, HHG in atoms and solids \cite{Watanabe1994,Vampa2015,VampaPRL2015,Orenstein2019,Dudovich2006,Mitra2020,Uzan2020}. It has also been shown that this scheme can be applied to study electron tunneling dynamics with attosecond time resolution \cite{Pedatzur2015,Dienstbier2023}. When combining the fundamental wave with its second harmonic frequency, the time symmetry of the waveform within one period is broken leading to emission of even order harmonic frequencies in centrosymmetric media. Even harmonic frequencies carry information about the quantum-mechanical phase acquired by the electron between the tunneling excitation and recombination. This phase is typically extracted via an interferometric measurement with the second harmonic field treated as a small perturbation, which does not shift the time window of electron tunneling with respect to the field of the fundamental wave \cite{Dudovich2006,Uzan2020}.

\begin{figure}
\includegraphics{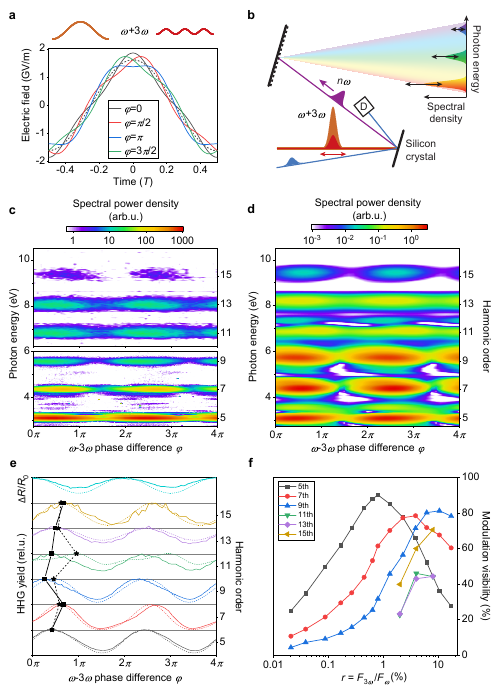}
\caption{(a) Time evolution of electric field of a combined $\omega$-3$\omega$ waveform for different values of the mutual phase difference $\varphi$ within one period $T$ of the fundamental wave (dashed curve). (b) Layout of the experimental setup. (c) Measured and (d) calculated dependence of high harmonic spectra generated in silicon on $\varphi$. The field ratios are $r=(3.9\pm0.5)\%$ in (c) and $r=4.3\%$ in (d). (e) Experimental (full curves) and numerically calculated (dashed curves) modulation of individual harmonic orders and the transient reflectivity signal of the probe pulse (top panel). The black points label the values of the mutual phase corresponding to the maxima of the measured (squares) and theoretical (stars) modulation for individual harmonics. (f) Measured visibility of modulation of the HHG yield for 5th-15th harmonics as a function of $r$.}
\label{fig1}
\end{figure}

In this report we demonstrate a scheme allowing to control the electron tunneling time within each half-cycle of the fundamental wave with attosecond precision using coherent superposition of an ultrashort infrared pulse with its third harmonic frequency. By controlling the mutual phase $\varphi$ between the $\omega$-3$\omega$ fields, the driving waveform changes (see Fig. \ref{fig1}a) leading to modulation of the amplitude and phase of high harmonic radiation generated in a silicon crystal. To separate the HHG process from electron tunneling excitation we simultaneously monitor the excited electron density as a function of the $\omega$-3$\omega$ phase by measuring the transient reflectivity of the sample after the interaction with $\omega$-3$\omega$ fields. We observe that even a small admixture of light at the third harmonic frequency to the fundamental pulse (the ratio between light intensities $I_{3\omega}/I_\omega \approx 10^{-4}-10^{-3}$) is sufficient to reach high modulation visibility of both the high harmonic generation yield and the excited electron population.

In our experiments we use the fundamental pulses in the mid-infrared spectral region with the photon energy of 0.62 eV and the duration of 35 fs. The third harmonic pulse (1.86 eV) is generated in a BBO crystal. Both pulses propagate collinearly and are overlapped in space and time at the surface of a silicon crystal (see the layout of the experimental setup in Fig. \ref{fig1}b, detailed description of the setup can be found in the Supplemental Material \cite{supp}). Both the $\omega$ and 3$\omega$ pulses have the same orientation of linear polarization along [100] crystallographic direction of silicon. The generated harmonic radiation is collected in the reflection geometry to avoid propagation effects in the sample. Due to the strong absorption of light at photon energies above the direct band gap (3.4 eV), the collected harmonic radiation is generated in a surface layer of the sample with the thickness of only $\approx$5-100 nm depending on the photon energy. The relative phase difference $\varphi$ between the $\omega$ and 3$\omega$ fields is controlled by a pair of fused silica wedges with a precision of 3 mrad corresponding to the time shift of the two waves by 3 as. (see Supplemental Material \cite{supp} for details). 


Simultaneously with the modulation of high harmonic yield we monitor the excited carrier population which remains in the sample after the interaction with $\omega$-3$\omega$ pulses using transient reflectivity of an ultraviolet probe pulse (3.62 eV, blue beam in Fig. \ref{fig1}b, details in Supplemental Fig. 3 \cite{supp}) incident on the sample with the delay time of 0.5 ps after the $\omega$-3$\omega$ pulse combination. This allows us to separate the modulation of the integrated tunneling rate from the modulation of the HHG yield. 

Coherent control of HHG in silicon by the two-color $\omega$-3$\omega$ field is experimentally demonstrated in Fig. \ref{fig1}c, where we plot the measured HHG spectra in the spectral region 2.8-10.5 eV as a function of the relative phase shift $\varphi$ between the fundamental (peak electric field in silicon 1.61 $\pm$ 0.1 GV/m) and the third harmonic fields with the ratio between the peak field amplitudes $r=F_{3\omega}/F_{\omega}=(3.9\pm 0.5)\%$. The results are compared with numerical calculations using time dependent density functional theory (TDDFT, details are described in Supplemental Material \cite{supp}) shown in Fig. \ref{fig1}d obtained with the fundamental electric field amplitude of 1.6 GV/m and the field ratio of $r=4.3\%$. We observe a clear modulation of the spectra both in the experimental and numerical data. The oscillations of individual harmonic orders obtained by integrating the power emitted in the spectral window of 0.1 eV around each harmonic peak are shown in Fig. \ref{fig1}e both for the experimental data (solid curves) and the numerical calculations (dashed curves). The data are normalized and vertically translated for clarity. 

The HHG modulation is compared to the modulation of the transient reflectivity of the sample $\Delta R/R_{\text{0}}$ shown as a solid curve in the uppermost panel of Fig. \ref{fig1}e. Here the dashed curve corresponds to the normalized population of electrons excited to the conduction band calculated by TDDFT (see Supplemental Material \cite{supp} for details). These data are used to obtain absolute calibration of the mutual phase between the $\omega$-3$\omega$ fields as the highest excited electron population is expected for $\varphi=0$, for which the maxima of both waveforms overlap (see Fig. \ref{fig1}a). We observe a good quantitative agreement between the experiment and numerical calculations both for the phase shifts of the oscillation maxima of individual harmonic orders, which are marked by squares (experimental data) and stars (theory), and for the depth of modulation of the high harmonic yield. 

The $\omega$-3$\omega$ phase difference of maximum generation yield differs for each harmonics as a consequence of the dispersion of the propagating electron and hole in higher energy bands of silicon, which tailors the interference between different quantum paths contributing to the interband emission of high energy photons. Remarkably, the maximum of the harmonic generation yield is reached for $\varphi \approx \pi /2$, which is shifted from the phase corresponding to the highest population of excited carriers. This observation suggests that the dominant mechanism of HHG driven by mid-infrared light in silicon is the interband polarization \cite{Suthar2022} in contrast to HHG in wide band gap materials excited by near-infrared light, where the intraband current was found to be dominating \cite{Garg2016}. The phase of HHG modulation maxima is only weakly dependent on the ratio between the amplitudes of electric field of the 3$\omega$ and $\omega$ pulses $r$ (see Supplemental Fig. 4 \cite{supp}). The modulation visibility of each harmonic frequency defined as $(S_\text{max}-S_\text{min})/(S_\text{max}+S_\text{min})$, where $S_\text{min}$ and $S_\text{max}$ are the minimum and maximum HHG yields at a particular harmonic frequency, is plotted in Fig. \ref{fig1}f as a function of $r$. The visibility reaches high values even for very weak third harmonic field only of about 1\% of the fundamental field.

To qualitatively understand the observed phenomena we recall the semi-classical model of HHG developed for atoms. In the adiabatic approximation, the instantaneous tunneling rate of an electron to the conduction band can be approximated using a Zener-like tunneling formula \cite{KANE1960,Keldysh1965}:

\begin{equation}
W(t) \propto \left | F(t) \right |^2 \exp \left (-\frac{\pi \sqrt{m^* E_\text{g}^3}}{2e\hbar \left | F(t) \right |} \right ).\label{eq1}  
\end{equation}

\begin{figure}
\includegraphics{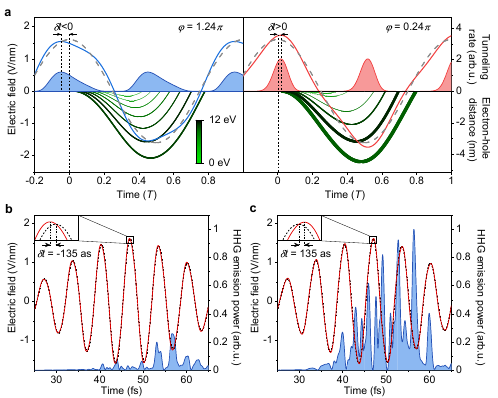}
\caption{(a) Semi-classical model showing the time evolution of the electric field of the combined $\omega$-3$\omega$ waveform for $\varphi=1.24\pi$ (blue curve) and $\varphi=0.24\pi$ (red curve). Instantaneous electron tunneling rate calculated using equation (\ref{eq1}) is shown as blue and red shaded areas. The classical trajectories of electron-hole pairs with reduced mass of $m^*=0.112m_0$ excited in different times are shown in the lower part of the figure with the color scale corresponding to the energy difference in the moment of recombination and the line thickness proportional to the tunneling probability associated with the particular trajectory. (b) The instantaneous power of high harmonic emission calculated using time-dependent density functional theory (blue shaded area) for $\varphi$=1.5$\pi$ and for the field ratio of $r$=4.3\%. The time evolution of the electric field of the combined waveform (red solid curve) is compared with the electric field of the fundamental pulse (black dashed curve). Inset: The peak of the waveform shifts by $\delta t$=-135 as. (c) The same as in (b) with $\varphi$=0.5$\pi$ leading to the positive time shift of the field maximum of $\delta t$=135 as.}
\label{fig2}
\end{figure}

Here $F(t)$ is the time dependent electric field in the material, $e$ is electron charge, $m^*$ is reduced mass, $E_\text{g}$ is band gap width and $\hbar$ is reduced Planck's constant. In Fig. \ref{fig2}a we plot the time evolution of the coherent superposition of the fundamental and third harmonic fields $F(t)=F_{\omega} \cos(\omega t)+F_{3\omega}\cos(3\omega t-\varphi)$ with the amplitudes $F_{\omega}$=1.6 V/nm and $F_{3\omega}$=0.16 V/nm (ratio $r=F_{3\omega}/F_{\omega}$=10\%) for two values of the relative phase difference $\varphi=1.24\pi$ (blue curve) and $\varphi=0.24\pi$ (red curve) corresponding to weak and strong emission of high harmonic radiation, respectively. The instantaneous tunneling rate describing the probability of electron transition to the conduction band per unit time calculated using Eq. (\ref{eq1}) with parameters $E_\text{g}=$3.4 eV (1st direct band gap of silicon) and $m^*=$0.112 $m_0$ is shown as blue and red shaded areas. The reduced mass corresponds to an electron-hole pair consisting of a light hole and an electron from the conduction band with low effective mass, which is energy degenerate with the lowest conduction band in the $\Gamma$ point of the Brillouin zone (see Supplemental Material \cite{supp} for details). We note that the time axis is defined with respect to the fundamental field oscillations. The classical trajectories $x(t)$ corresponding to the distance between electron and hole are calculated by solving the equation of motion $\ddot{x}=-eE/m^*$ in parabolic approximation. The trajectories corresponding to different tunneling times within one half-cycle of the fundamental wave are shown as curves in the lower section of Fig. \ref{fig2}a. The color scale corresponds to the total energy of the recombining electron and hole $E_\text{tot}=1/2m^* \dot{x}^2+E_\text{g}$ while the line thickness is proportional to the tunneling probability of the electron corresponding to the particular trajectory.

There are two important effects which contribute to the modulation of the HHG yield. The first effect is the modulation of the instantaneous tunneling rate induced by the combined $\omega$-3$\omega$ waveform. The second effect contributing to the HHG modulation is the time shift $\delta t$ of the maximum of the tunneling window. The time shift has important implications for the probability of electron-hole recombination. When HHG is driven by a single frequency field, only the electrons generated during the second half of each tunneling window can recombine and emit photons while the ones created in the first half of the tunneling window propagate in the laser field away from the original position. The time shift of the tunneling window to later times with respect to the fundamental wave thus increases the number of recombining electron-hole pairs contributing to high harmonic emission while the shift to earlier times leads to the opposite effect. This model suggests that the maximum of the high harmonic yield is reached for different value of $\varphi$ than the maximum of the excited carrier population, which agrees with our experimental observations shown in Fig. \ref{fig1}e. 


While this simple model captures qualitatively the features observed in the experiments, it cannot describe the complex electron dynamics in silicon involving the intraband currents and interband transitions between multiple bands. To get a deeper insight into the coherent highly nonlinear response of silicon we analyse the nonlinear current obtained from TDDFT simulations. In Figs. \ref{fig2}b,c we plot the combined $\omega$-3$\omega$ waveforms with $r$=5\% for two different values of $\varphi$=1.5$\pi$ (red curve in Fig. \ref{fig2}b) and $\varphi$=0.5$\pi$ (red curve in Fig. \ref{fig2}c) along with the calculated instantaneous power of the emitted high harmonic radiation at photon energies above 2.8 eV (blue shaded areas). The field of the fundamental pulse is shown for comparison as dashed curves in Figs. \ref{fig2}b,c. The time shifts of the maxima of the waveforms in these two cases of $\delta t=-135$ as (see the inset of Fig. \ref{fig2}b) and $\delta t=135$ as (see the inset of Fig. \ref{fig2}c), respectively, lead to strong modulation of the emission probability even with a weak perturbation of the fundamental waveform by the third harmonic field. We observe that the maxima and minima of the HHG yield correspond to approximately the same values of excited electron population, emphasizing the role of the time shift of the electron tunneling window which is evidenced to be the dominant mechanism causing the observed modulation of HHG in the silicon crystal.


When the time of electron excitation shifts with respect to the oscillation maxima of the fundamental wave, not only the amplitude but also the phase of the generated harmonic radiation is expected to shift. To measure the phase delays of the emitted high harmonic radiation we apply spectral interferometry by using a signal and reference high harmonic fields \cite{Lu2019,uchida2023high}. The reference field is generated by an infrared pulse phase-locked to the $\omega$-3$\omega$ combination, which arrives to the sample before the two-color waveform (layout of the experimental setup is shown in Fig. \ref{fig3}a, the details are described in Supplemental Material \cite{supp}). The time separation of the signal and reference fields of about 170 fs leads to spectral interference in the harmonic spectra, example of which is shown in Fig. \ref{fig3}b for 5th to 9th harmonic frequencies. By monitoring the shifts of the spectral interference fringes we measure the relative shifts of the emission phase of harmonic radiation as a function of $\varphi$. The measured phase shift of harmonic emission $\varphi_\text{HHG}$ corresponds to the emission time delay of $\delta t_\text{em}=\varphi_\text{HHG}/\omega_\text{HHG}$. We note that the carrier population generated in the sample during the HHG by the reference pulse is not expected to significantly influence the phase of the coherent photons generated by the signal pulse. 


\begin{figure}
\includegraphics{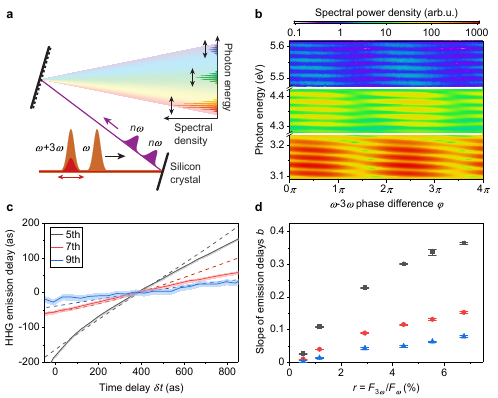}
\caption{(a) Layout of the experimental setup for spectral interferometry of high harmonic fields. (b) Spectral interference of the 5th, 7th and 9th harmonic frequencies measured as a function of $\varphi$. (c) Measured (solid curves, $r$=5.5\%) and calculated (dashed curves, $r$=5\%) attosecond emission delays obtained from the data shown in (b) as a function of the time delay of the 3$\omega$ field with respect to the $\omega$ field. (d) Slopes of the measured emission delays of the 5th (black squares), 7th (red circles) and 9th (blue triangles) harmonics as a function of the field ratio $r$ obtained by fitting the data shown in (c) using linear function in the time interval of 400 as around the modulation maxima.}
\label{fig3}
\end{figure}

In Fig. \ref{fig3}c we show the measured high harmonic emission delays (field ratio $r$=5.5\%) compared to the delays obtained from the numerical TDDFT simulations ($r$=5\%) in the time window around the maxima of the harmonic emission shown in Fig. \ref{fig3}b. We observe that close to the maximum of the emission yield, the emission delay scales approximately linearly with the time delay between the two-color fields $\delta t$ with different slopes for individual harmonic frequencies. The emission delays of high harmonic photons relative to the fundamental wave thus depend on the photon energy and cannot be understood only by considering the time shift of the electron tunneling time. In contrast, the emission phase is determined by the combination of the tunneling time and the quantum mechanical phase acquired by the electron-hole wavepacket during its coherent dynamics between tunneling and recombination, where multiple bands are involved in the high harmonic generation process \cite{Kartner2019,Suthar2022}. The measured slopes $b$ of the emission delays of 5th to 9th harmonics obtained by fitting the data shown in Fig. \ref{fig3}c by a linear function $y=a+bx$ are plotted in Fig. \ref{fig3}d as a function of the $\omega$-3$\omega$ field ratio $r$. 

The observed smaller delays of higher order harmonics can be qualitatively understood in the framework of the semi-classical electron dynamics shown in Fig. \ref{fig2}a. The electrons propagating along trajectories corresponding to the highest recombination energy are generated only in a short time window within one half-cycle of the fundamental wave. The recombination time of these high energy electrons practically does not shift with the relative phase $\varphi$. However, the electrons with lower recombination energies are generated in two distinct time windows corresponding to short and long trajectories. Photons within the intermediate energy range are thus emitted in two distinct times within each half-cycle of the $\omega$ field (dark green curves in Fig. \ref{fig2}a) with different amplitudes and emission phases. The change of the relative phase shift $\varphi$ between the $\omega$-3$\omega$ field leads to strong changes of the recombination probability of particular trajectories, which in turn causes the larger emission delays.


In summary, the combination of coherent two-color control and phase-resolved detection scheme with attosecond resolution introduced here brings opportunities for gaining a deeper understanding of HHG in solids. In particular we study the relation between the tunneling time, which is controlled within each half-cycle of the driving wave by adding a weak perturbation at the third harmonic frequency, and the amplitude and phase of the emitted high-energy photons. A direct comparison between the modulation of high harmonic emission and the modulation of the real carrier population induced by the $\omega$-3$\omega$ fields allows to separate the processes of electron excitation from the nonlinear coherent dynamics leading to HHG. We note that the coherent electron-hole pair dynamics can differ considerably in systems with strong many-body correlations (e.g. 2D crystals), which have recently been studied using a combination of ultrashort infrared excitation pulse with oscillating THz field driving electron-hole recollision paths \cite{Freudenstein2022}. Further, the $\omega$-3$\omega$ combination may find applications in atomic HHG for enhancement of the emission yield or for advanced gating techniques combining the coherent control with dynamic phase-matching \cite{Thomann2009}.

\begin{acknowledgments}
\paragraph{Acknowledgments}
Czech Science Foundation (project GA23-06369S), Charles University (UNCE/SCI/010, SVV-2020-260590, PRIMUS/19/SCI/05, GAUK 349921 and 124324). Funded by the European Union (ERC, eWaveShaper, 101039339). Views and opinions expressed are however those of the author(s) only and do not necessarily reflect those of the European Union or the European Research Council Executive Agency. Neither the European Union nor the granting authority can be held responsible for them. T. J.-Y. D. acknowledges support from the European Regional Development Fund and the State Budget of the Czech Republic (Project SENDISO No.
CZ.02.01.01/00/22\_008/0004596), from the Ministry of Education, Youth and Sports of the Czech Republic through the e-INFRA CZ (ID:90254). This work was supported by TERAFIT project No. CZ.02.01.01/00/22\_008/0004594 funded by OP JAK, call Excellent Research.

\end{acknowledgments}

\textit{Data availability} - The data that support the findings of this article are openly available at Zenodo \cite{Data}.

\nocite{*}

\providecommand{\noopsort}[1]{}\providecommand{\singleletter}[1]{#1}%

\end{document}



\title{Attosecond control of solid-state high harmonic\\
generation using $\omega$-$3\omega$ fields - Supplemental Material}


\author{Adam Gindl}
    \affiliation{Faculty of Mathematics and Physics, Charles University, Ke Karlovu 3, 12116 Prague 2, Czech Republic}
\author{Pawan Suthar}
    \affiliation{Faculty of Mathematics and Physics, Charles University, Ke Karlovu 3, 12116 Prague 2, Czech Republic}
\author{František Trojánek}
    \affiliation{Faculty of Mathematics and Physics, Charles University, Ke Karlovu 3, 12116 Prague 2, Czech Republic}
\author{Petr Malý}
    \affiliation{Faculty of Mathematics and Physics, Charles University, Ke Karlovu 3, 12116 Prague 2, Czech Republic}
\author{Thibault J.-Y. Derrien}
    \affiliation{Institute of Physics of the Czech Academy of Sciences, Na Slovance 1999/2, 18200 Prague, Czech Republic}
    \affiliation{IT4Innovations, Technical University of Ostrava, 17. listopadu 2172/15, 70800 Ostrava-Poruba, Czech Republic} 
\author{Martin Kozák}
    \email{m.kozak@matfyz.cuni.cz}
\affiliation{Faculty of Mathematics and Physics, Charles University, Ke Karlovu 3, 12116 Prague 2, Czech Republic}

\maketitle

\tableofcontents

\section{Experimental setup}

The experiments demonstrating coherent two-color control of high harmonic generation in silicon are performed using pulses generated in a noncollinear optical parametric amplifier with subsequent difference frequency generation (NOPA-DFG, setup is described in detail in \cite{Kozak2021}) and the pulses at its third harmonic frequency. The NOPA-DFG setup is pumped by solid-state femtosecond laser system Pharos SP2-6W (Light  Conversion) with ytterbium-doped active medium. The NOPA-DFG output pulses have central wavelength of 2000 nm and pulse duration at the sample of 35 fs (FWHM of instantaneous power). The coherent $\omega$-3$\omega$ pulse combination is generated in the setup, which is shown in Supplementary Fig. S\ref{figS1}. The beam passes through a half wave plate to control the direction of linear polarization of the fundamental pulse. Subsequently, the beam is focused using an off-axis parabolic mirror with a focal distance of 15 cm to a BBO crystal, where the pulses at third harmonic frequency 3$\omega$ are generated. We use phase-matching angle of $\theta$=27.2° for direct third harmonic generation ($oooe$-type phase-matching) leading to vertically polarized 3$\omega$ pulse (polarization along the direction of projection of the optical axis of the BBO crystal in the plane perpendicular to the light propagation). In the first experiment, the half wave plate is adjusted such that the fundamental polarization is horizontal. The pulse thus propagates as an ordinary ray and we obtain only one replica of the fundamental pulse after the BBO. When the half wave plate is rotated to 22.5°, the polarization incident to BBO crystal has both the horizontal and vertical components with the same amplitude. As a consequence of the birefringence of the BBO crystal we obtain two orthogonally polarized pulses, which are delayed due to different group indexes of the ordinary and the extraordinary rays. After the BBO crystal, the two collinear beams at the fundamental and the third harmonic frequencies are collimated by a second off-axis parabolic mirror with focal length of 15 cm.

The relative phase between the $\omega$ and 3$\omega$ pulses $\varphi$ is controlled using a pair of fused silica wedges with an apex angle of $\alpha$=4°. One of the wedges is placed on a translation stage. The phase shift of the third harmonic field with respect to the fundamental pulse resulting from the propagation in the wedges over a distance $\Delta L$ can be expressed as $\varphi=3 \omega \Delta L \left( n_{3\omega}-n_{\omega} \right)/c$ , where $c$ is speed of light in vacuum and $n_{\omega}$=1.4381 and $n_{3\omega}$=1.4561 are the refractive indexes of fused silica at frequencies $\omega$ and $3\omega$, respectively. One period of the high harmonic yield modulation corresponds to the shift of the relative $\omega$-$3\omega$ phase $\varphi$ by 2$\pi$ which gives $\Delta L=$37 $\mu$m. The calculated shift of the wedge needed to change the thickness of the material by $\Delta L$ is $\Delta x=\Delta L/\tan(\alpha)=0.59$ mm, which matches well the experimentally determined value of 0.574$\pm$0.005 mm. While the refractive indexes of fused silica at frequencies $\omega$ and 3$\omega$ differ strongly, the difference between group velocities of the two pulses in the wedges are much smaller (group indexes are $n_{g,\omega}$=1.4673 and $n_{g,3\omega}$=1.4732) leading to a small relative shift of the pulse envelopes when changing the relative phase by several periods of the $3\omega$ field. To compensate the total group delay which the $\omega$ and 3$\omega$ pulses obtain between the first BBO crystal and the sample we let the beam pass through a second BBO crystal with the optical axis cut under the angle $\theta$=50°. Due to the negative uniaxial birefringence of BBO, the ordinary ray at frequency $\omega$ propagates slower in the crystal than the extraordinary ray at 3$\omega$. The beam is finally focused on the sample surface by an off-axis (90°) silver coated parabolic mirror with focal length of $f$=35 mm. The spot sizes of the two beams on the sample are $w_\omega=15$ ${\mu}$m and $w_{3\omega}=8$ ${\mu}$m. The dependence of nonperturbative HHG yield in silicon on the incident light intensity of the fundamental pulses was previously measured to be cubic \cite{Suthar2022}. When considering Gaussian profile of the fundamental beam, the harmonics are produced from the area with a radius decreased by a factor of $w_{HHG}=w_\omega / \sqrt{3}=8.7$ ${\mu}$m, which is well matched to the size of the third harmonic beam. To adjust the polarizations of all the pulses to be linear in the same direction we put a thin polarizer (Thorlabs, LPNIRA) between the final focusing parabolic mirror and the sample.

High harmonic radiation is collimated by a UV fused silica lens with focal distance of 100 mm. The spectra are detected using a grating spectrograph (Andor, Shamrock 163) with a grating containing 600 lines/mm blazed for 300 nm. The spectrometer is equipped with a cooled CCD camera (Andor iDUS 420). Prior to entering the spectrometer, the light is spectrally dispersed in the vertical direction by a pair of fused silica prisms. This allows to filter out most of the radiation at wavelengths above 500 nm before the high harmonics enter the spectrometer. Because the dispersion is applied in the vertical direction, all the harmonics are transmitted through the entrance vertical slit of the spectrometer. The harmonics 11th-15th have photon energies higher than 6 eV and are strongly absorbed in air. The modulation of the high harmonic orders is thus measured in vacuum using a vacuum ultraviolet spectrometer (EasyLight, H+P Spectroscopy) combined with a microchannel plate detector.

Due to the inline setup of the nonlinear interferometer, in which both beams propagate through the same beam paths, the drifts of the relative phase difference $\varphi$ between the $\omega$ and 3$\omega$ pulses are negligible and the time jitter of the two fields is only few attoseconds over the measurement time of one hour (see the measured modulation of the 5th harmonic frequency over the time period of 60 minutes shown in Supplementary Fig. S\ref{figS2}).

The carrier population excited in the sample by the coherent superposition of $\omega$-3$\omega$ fields is monitored by measuring the transient reflectivity of the sample. We use an independent probe pulse with the wavelength of 343 nm and photon energy of 3.62 eV (third harmonics of the Pharos laser output at 1030 nm), which is incident on the sample with a controlled time delay with respect to the coherent $\omega$-3$\omega$ pulse combination. The high photon energy of the probe is selected due its short penetration depth into silicon of only few nanometers, which prevents accumulation of the signal coming from different depths which would be modified by the fact that the phase velocities of the pulses at $\omega$ and 3$\omega$ frequencies in silicon strongly differ. The second reason for selecting a short probe wavelength is the possibility to focus the probe beam to a smaller spot size than the $\omega$ and 3$\omega$ beams, which allows to probe the region of the sample with approximately homogeneous spatial distribution of the excited carriers. In this experiment, the pump $\omega$-3$\omega$ combination propagates through an optical chopper, which modulates the beam. After being reflected from the sample, the power of the probe beam at 3.62 eV is detected by a silicon photodiode. The electronic signal is measured using a lock-in amplifier (SR830, Stanford Research Systems) at the frequency of the optical chopper. To verify that the transient reflectivity signal at 3.62 eV scales linearly with the excited carrier population we perform additional degenerate pump-probe experiment, in which we use the same photon energy to excite the electron-hole pairs in silicon. The measured transient reflectivity dynamics at probe photon energy of 3.62 eV is shown in Supplementary Fig. S\ref{figS3} as a function of the time delay between the pump and probe pulses. We observe that the magnitude of the transient reflectivity signal in the time delay of 500 fs (the same time delay was used for probing the excited carrier population with $\omega$-3$\omega$ pump) scales linearly with pump fluence (see the inset of Supplementary Fig. S\ref{figS3}). 
	
	
We note that the physical origin of the transient reflectivity signal at probe photon energy of 3.62 eV is rather indirect. After the excitation over the direct band gap of silicon with the width of about 3.4 eV, the electrons and holes relax rapidly to the minima of the respective bands via electron-phonon scattering. For this reason, absorption bleaching of the direct transition is expected to have very fast dynamics. Similarly, the signal corresponding to excited carrier (Drude-like) absorption is expected to be weak since it scales with $\lambda^2$. For these reasons we conclude that the most probable origin of the signal is the increase of the lattice temperature after the initial relaxation of the excited carrier system, which transfers the excess energy to the lattice vibrations. The peak excited carrier density in our experiments can be estimated as $n=\alpha_\text{pump} (1-R_\text{pump})f_\text{pump}/(\hbar \omega_\text{pump})$, where $\alpha_\text{pump}=1.1\times10^6$ cm$^{-1}$ is the absorption coefficient and $R_\text{pump}=0.56$ is reflectivity of silicon at photon energy of 3.62 eV, $f_\text{pump}$ is the fluence of the pump pulse and $\hbar \omega_\text{pump}=3.62$ eV is the pump photon energy. When we consider the specific heat capacity of silicon at room temperature $c_p=0.7$ J/(g.K), the excess energy per electron-hole pair of $\Delta E=2.5$ eV and the density of silicon $\rho=2.33$ g/cm$^{-3}$, the estimated increase of crystal temperature at excited carrier density of $n=10^{21}$ cm$^{-3}$ is $\Delta T=n\Delta E/(\rho c_p)=245$ K. Both the real and the imaginary parts of dielectric function of silicon decrease with increasing lattice temperature in this spectral region \cite{Sik1998}, which leads to a decrease of the reflectivity of the sample in agreement with our observations. 

\begin{figure}
	\includegraphics[width = 0.7\linewidth]{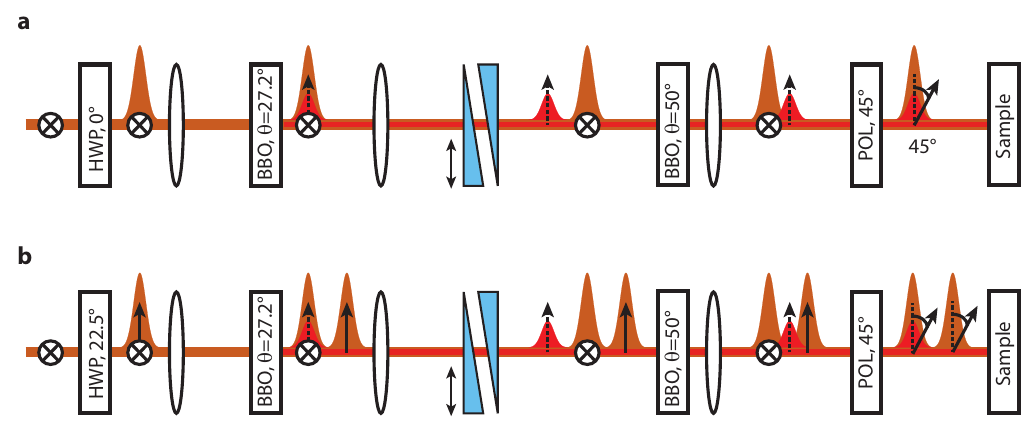}
	\caption{Experimental setup for generation of coherent superposition of $\omega$-3$\omega$ pulses. (a) The setup for the measurements of amplitude modulation of high harmonic generation. The half wave plate (HWP) is rotated to 0° (fundamental polarization remains horizontal). The pulse is focused by off-axis parabolic mirror to the first beta-barrium borate (BBO) crystal phase-matched for direct third harmonic generation. After the crystal, the beam is collimated and propagates through the fused silica wedges, where the relative $\omega$-3$\omega$ phase is adjusted by translating one of the wedges. The group delay between the pulses is compensated in the second birefringent BBO crystal. The beam is focused on the sample by an off-axis parabolic mirror (all mirrors are displayed as lenses for simplicity). Both pulses propagate through a thin polarizer rotated under 45° to generate linear polarizations with their axis being parallel to each other on the sample surface. (b) The setup for the measurements of attosecond phase delays of high harmonic radiation. The components are the same as in (a) with the half wave plate rotated by 22.5° leading to polarization rotation of the fundamental beam by 45°. As a consequence, two temporally separated fundamental pulses leave the first BBO crystal and co-propagate to the sample, where the latter pulse is temporally overlapped with the 3$\omega$ pulse and the former one serves as a reference pulse for the spectral interferometry measurements.}
	\label{figS1}
\end{figure}

\begin{figure}
	\includegraphics[width = 0.8\linewidth]{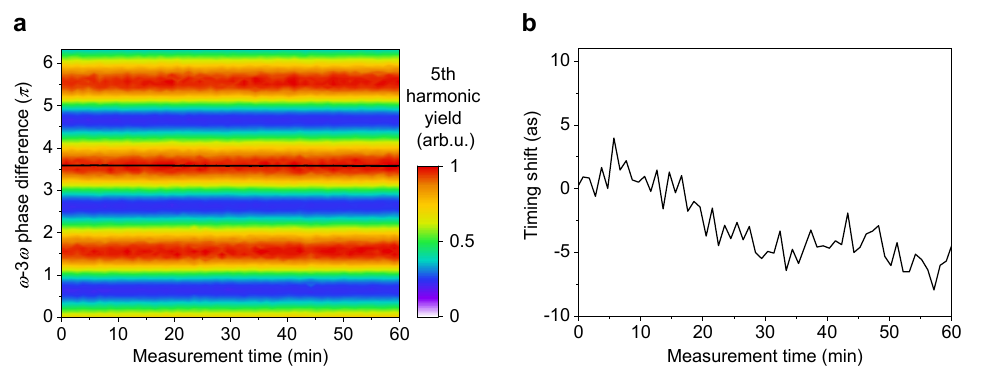}
	\caption{Long-term stability of $\omega$-3$\omega$ interferometer. (a) Measurements of the modulation of 5th harmonic generation as a function of the relative phase difference between $\omega$-3$\omega$ fields $\varphi$ repeated over 60 minutes (color scale) along with the phase of the maxima obtained from fitting the data with harmonic function (black curve). (b) Timing jitter and long-term drift calculated from the measured data shown in (a). The RMS timing fluctuations of the $\omega$-3$\omega$ fields measured over 60 minutes are 2.8 as.}
	\label{figS2}
\end{figure}

\begin{figure}
	\includegraphics[width = 0.4\linewidth]{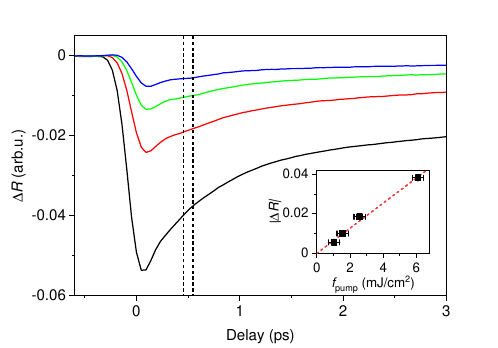}
	\caption{Transient reflectivity dynamics in silicon. The curves show the dynamics of the transient reflectivity change induced by the pump pulse with the photon energy of 3.62 eV probed at the same photon energy. The curves are obtained with the pump fluence of 6.1 mJ/cm$^2$ (black), 2.6 mJ/cm$^2$ (red), 1.56 mJ/cm$^2$ (green) and 1.02 mJ/cm$^2$ (blue). The inset shows the dependence of the magnitude of the transient reflectivity signal $|\Delta R|$ as a function of the pump fluence (squares) compared to the linear fit (dashed line).}
	\label{figS3}
\end{figure}

\clearpage

\section{Dependence of HHG modulation on the ratio $r$ between the field amplitudes of $\omega$-3$\omega$ pulses}

The modulation of the 5th, 7th and 9th harmonic frequencies was measured as a function of the ratio $r=F_{3\omega}/F_{\omega}$ between the amplitudes of electric field of the pulse at frequency $3\omega$ and the pulse at frequency $\omega$. The results are plotted in Supplementary Figure S\ref{figS4}. The black squares mark the mutual phase $\varphi$ of the modulation maxima for each harmonic frequency. We observe that the value of $\varphi$ corresponding to the modulation maxima changes only slightly over a large range of field ratios.

\begin{figure}[H]
	\centering
	\includegraphics[width = 0.5\linewidth]{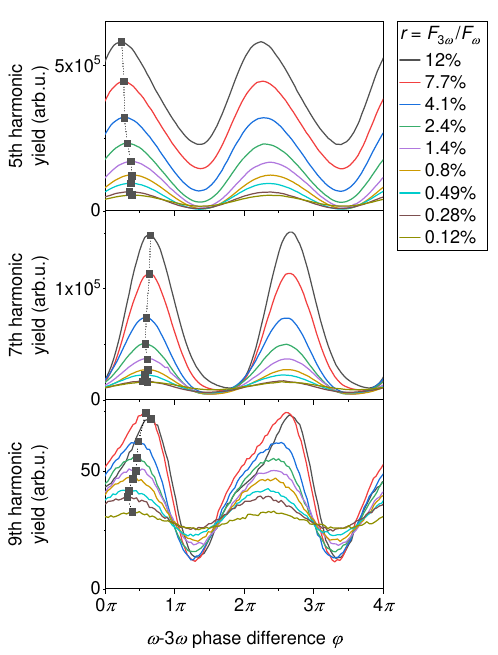}
	\caption{Modulation of high harmonic generation yield in silicon by $\omega$-3$\omega$ fields for different field ratios $r$. Curves show the measured data of the HHG yield of the 5th harmonics (uppermost panel), 7th harmonics (central panel) and 9th harmonics (bottom panel) as a function of the relative phase between $\omega$-3$\omega$ fields $\varphi$. The phase corresponding to the maxima of modulation for each value of $r$ are marked by black squares.}
	\label{figS4}
\end{figure}

\clearpage

\section{Data processing}

The measured high harmonic spectra are not compensated for the absolute spectral efficiency of the detection setup. The absolute calibration is not required because we are only interested in the relative changes of the HHG yield of individual harmonic frequencies as a function of experimental parameters (mutual phase difference between $\omega$-3$\omega$ fields, ratio of the field amplitudes $r$). The generation yield at each harmonic frequency is obtained by integrating the spectral window with the width of 0.1 eV around the spectral peak of the particular harmonics.

The spectral interferometry data shown in Fig. 3b are processed by fitting the peaks of the individual harmonics in each spectrum corresponding to specific $\omega$-3$\omega$ phase differences $\varphi$ using a function:
\begin{equation}
\label{eq:SpectInterFit}
{f(\hbar \omega)=A_1 \text{exp} \left [ -\frac{(\hbar \omega-\hbar \omega_{HHG})^2}{2\Delta\omega_{HHG}^2} \right ]+A_2 \text{exp} \left [ -\frac{(\hbar \omega-\hbar \omega_{HHG})^2}{2\Delta\omega_{HHG}^2} \right ]\text{sin} \left( \omega \tau + \Delta \varphi_{HHG} \right) + A_3}
\end{equation}
Here $\omega_{HHG}$ and $\Delta\omega_{HHG}$ are the central frequency and the spectral width of the particular harmonic peak. Fitting parameters are the delay between the two harmonic pulses $\tau$, their relative phase shift $\Delta \varphi_{HHG}$ and the coefficients $A_1$, $A_2$ and $A_3$ which correspond to the amplitude of the slowly varying Gaussian envelope ($A_1$), the amplitude of the spectral interference ($A_2$) and the constant background ($A_3$). The data are fitted in two steps. First the time delay $\tau$ is determined from the data with the strongest spectral interference shown in Supplementary Fig. S\ref{figS5}. In the second step, $\tau$ is kept constant and the relative phase shift of harmonic fields $\Delta \varphi_{HHG}$ is obtained for different values of the relative phase between the driving $\omega$-3$\omega$ fields. The relative emission delays shown in Fig. 3c are calculated from the relative phase shifts of the individual harmonic peaks as $\delta t_{em}=\Delta \varphi_{HHG}/\omega_{HHG}$, where $\omega_{HHG}$ is the central frequency of the particular harmonics.

\begin{figure}[H]
	\centering
	\includegraphics[width = 0.6\linewidth]{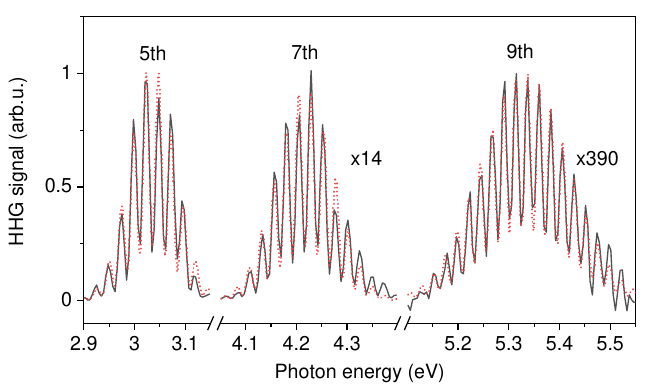}
	\caption{Spectral interferometry of HHG in silicon. The measured data (black curve) of the spectral interference of the 5th, 7th and 9th harmonics are fitted using Eq. (\ref{eq:SpectInterFit}) (red dotted curves).}
	\label{figS5}
\end{figure}

\section{Details of the semi-classical model}

In the semi-classical model we assume that the electron-hole pair is excited by quantum tunneling with the rate given by Eq. (1) of the main manuscript. The classical trajectories corresponding to the relative electron-hole distance in the electric field of the combined $\omega$-3$\omega$ waveform are calculated in parabolic approximation. We assume that the highest contribution comes from the electron-hole pairs excited in the $\Gamma$ point of the Brillouin zone in the bands with the lowest value of the reduced mass. This assumption is based on the fact that the tunneling rate resulting from Eq. (1) of the main manuscript increases with the decreasing distance between the conduction and valence bands and with the decreasing reduced mass of electron-hole pair. The band structure of silicon in $\Gamma$-X direction contains two conduction and two valence bands, which are energy degenerate in the $\Gamma$ point. Dispersion of these bands is calculated using density functional theory and the results are shown in Supplementary Fig. S\ref{figS6}. The lowest reduced mass in the $\Gamma$ point of $m^*=$0.112 $m_0$ corresponds to a combination of an electron in the red band and a hole in the blue band.

\begin{figure}[H]
	\centering
	\includegraphics[width = 0.5\linewidth]{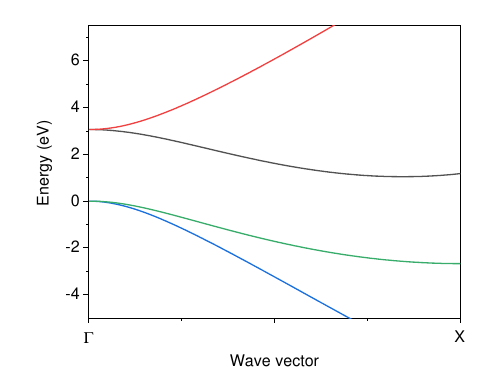}
	\caption{Two lowest conduction bands and two highest valence bands of silicon. In the semi-classical model we assume a quasiparticle with the reduced mass of $m^*=$0.112 $m_0$ corresponding to an electron in the red band and a hole in the blue band in the $\Gamma$ point. }
	\label{figS6}
\end{figure}

\section{Numerical calculations using time dependent density functional theory (TDDFT)}

\subsection{Ground state preparation}

Silicon (100) crystal is prepared using a primitive cell made of two silicon
atoms along with non-orthogonal periodic boundary conditions
as implemented in \emph{Octopus} \cite{Tancogne-Dejean2019c}. Inter-atomic distance is
chosen equal to the experimental value ($a=5.431$ angstroms) \cite{Okada1984}.
The inter-atomic potential is described using the method of norm-conserving
\emph{ab-initio} pseudo-potentials \cite{Troullier1991}. The real-space
is meshed regularly using a step of $\delta x=0.23$ angstroms, and
the momentum space is meshed uniformly using a grid of $k=40\times40\times40$,
repeated 4 times centered at points X and L. 

The ground state is prepared using the Tran-Blaha functional (TB09)
\cite{Tran2009}. The convergence of the ground state was verified
by comparison with available literature of more precise methods \cite{Waroquiers2013}.
The resulting direct band gap $E_{\Gamma}$ of silicon is found to be $E_\text{g}$=3.04 eV. 

\subsection{Time-dependent simulations and control of convergence}

The time-evolved Kohn-Sham (KS) states are prepared by solving the
KS equation expressed in the velocity gauge and in Hartree atomic
units:
\begin{gather}
\left[\left(-\frac{i\hbar}{2m_{e}}\boldsymbol{{\nabla}_{r}}+\frac{\left|e\right|}{c}\boldsymbol{A}\left(t\right)\right)^{2}+\hat{v}_{\text{ion}}\left(\boldsymbol{r}\right)+\hat{v}_{\text{H}}\left[n\left(\boldsymbol{r},t\right)\right]\left(\boldsymbol{r}\right)+\right.\label{eq:KSequation}\\
\left.+\hat{v}_{\text{xc}}\left[n\left(\boldsymbol{r},t\right)\right]\left(\boldsymbol{r}\right)\right]\times\psi_{n,\boldsymbol{k}}\left(\boldsymbol{r},t\right)=i\hbar\frac{\partial}{\partial t}\psi_{n,\boldsymbol{k}}\left(\boldsymbol{r},t\right).\nonumber 
\end{gather}
Here the vector potential describing the laser pulse $A(t)$ is introduced
using the dipolar approximation in the minimal coupling formulation.
In Eq. (\ref{eq:KSequation}), $i$ is the pure imaginary number,
$\hbar$ is the reduced Planck constant, $m_{e}$ is the electron
mass at rest, $\boldsymbol{{\nabla}_{r}}$ is the real-space
gradient operator, $\left|e\right|$ is the elementary charge of the
electron and $c$ is the light velocity. In the Coulomb gauge, the
vector potential $\boldsymbol{A}\left(t\right)$ in the dipolar approximation
is related to the electric field $\boldsymbol{F}\left(t\right)$ via
the relation
\begin{equation}
\boldsymbol{A}\left(t\right)=-c\int_{-\infty}^{t}\boldsymbol{F}\left(t'\right)\,dt'.
\end{equation}

In Eq. (\ref{eq:KSequation}), $\hat{v}_{\text{ion}}\left(\boldsymbol{r}\right)$
represents the potential of the atomic lattice. Note that the non-local
term originating from the pseudo-potential is not detailed for simplicity.
$\hat{v}_{\text{H}}$ denotes the Hartree potential, corrected by
the exchange-correlation potential noted $\hat{v}_{\text{xc}}$. The
incident wavelength is normalized to the band-gap energy obtained
using the TB09 functional. Orientation of the linearly-polarized electric
field is set along the $(100)$ direction of the cubic crystal. From the time-evolved Kohn-Sham orbitals $\psi_{n,\boldsymbol{k}}\left(\boldsymbol{r},t\right)$,
the time-dependent electrical current $\boldsymbol{J}\left(t\right)$
is computed as indicated in Ref. \cite{Wachter2014}. 

The electric field of laser pulses at frequencies $\omega$ and 3$\omega$, respectively, which is used in the numerical simulations can be written as:
\begin{equation}
\vec{F}(t)=\vec{F}_{\omega}\times\text{sin}\left(\omega t+\varphi_{0,\omega}\right)\times f\left(t\right)+\vec{F}_{3\omega}\times\text{sin}\left(3\omega t+\varphi_{0,3\omega}-\varphi\right)\times f\left(t\right),\label{eq:FieldExperimental}
\end{equation}
where: 
\begin{equation}
f\left(t\right)=\text{\ensuremath{\sin^{4}}\ensuremath{\left(\frac{\pi}{2}\frac{t-t_{0}-\tau_{\text{corr}}}{\tau_{\text{corr}}}\right)}}\\
\times H\left(t-t_{0}+\tau_{\text{corr}}\right)\left[1-H\left(t-t_{0}-\tau_{\text{corr}}\right)\right].
\end{equation}
Here the correction of the FWHM for $\sin^4$ envelope can be expressed as: 
\begin{equation}
\tau_{\text{corr}}=\frac{4\tau_{\text{exp}}}{\pi}\arcsin\left(2^{-1/4}\right).
\end{equation}
 $\tau_{\text{exp}}$=35 fs is the experimental pulse duration and $H(x)$ is Heaviside function of argument $x$. Note
that 4th-power sinus is commonly employed to decrease the spurious
oscillations originating from the Fourier transformation upon computation
of high-harmonic spectrum \cite{Sato2021,Suthar2022}.

\subsection{Normalization of pulse frequency to band gap energy}

Density functional theory gives for most of the materials systematically lower values of the band gap compared to the experimental values. For the HHG process, an important parameter is the ratio between the driving photon energy and the minimum band gap. As shown in \cite{Suthar2022}, the driving frequency can be normalized to the reduced band-gap provided by DFT \cite{Waroquiers2013} to fulfill the relation $
{\hbar \omega_{\text{TB09}}}/{E_{\text{g},\text{TB09}}} = {\hbar \omega_{\text{exp}}}/{E_{\text{g},\text{exp}}}$. In the calculations, the wavelengths of $\lambda_{\omega,\text{TB09}}$=2237 nm and $\lambda_{3\omega,\text{TB09}}$=745.66 nm are used instead of the experimental values 2000 nm and 666.67 nm, respectively. The time axis in the calculation results is normalized back to obtain harmonic spectra with frequencies corresponding to experimental data.

\subsection{Calculation of harmonic spectra}

Harmonic spectra are calculated using Larmor's formula and Fourier transform of the total current obtained by the TDDFT simulations. The Fourier transform is applied to a temporal window that decays in $\sin^4$ at the end of pulse to avoid generating spurious oscillations in the spectra as a consequence of no dephasing present in the TDDFT calculations (similarly to Ref. \cite{Suthar2022}). 

\subsection{Calculation of the excited electron density}

The number of electrons excited from the valence band to the conduction bands is computed using the method described in Ref. \cite{Derrien2021}. Note that transient values of electron density are subjected to gauge dependencies \cite{Ernotte2018}. Therefore, reported values of quantity of excited electrons are captured after the laser pulse, i.e., in absence of fields, a situation where gauge is not problematic \cite{Ernotte2018,Otobe2008}.

\nocite{*}

\providecommand{\noopsort}[1]{}\providecommand{\singleletter}[1]{#1}%